\documentstyle[12pt,aasms4]{article}
\def\kms{km~s$^{-1}$}

\def\cm2{cm$^{-2}$}

\def\lya{Ly$\alpha$ }

\slugcomment{To appear on AJ, April 1998}

\lefthead{Fontana et al.}
\righthead{Star--formation at $z=4.7$}

\begin{document}
\newcommand{\lsim}{\ \raise -
2.truept\hbox{\rlap{\hbox{$\sim$}}\raise5.truept
	\hbox{$<$}\ }}			
\newcommand{\gsim}{\ \raise -
2.truept\hbox{\rlap{\hbox{$\sim$}}\raise5.truept
 	\hbox{$>$}\ }}		

\title{Star--formation at $z=4.7$ in the environment of the QSO BR1202-07
\altaffilmark{1}}

\altaffiltext{1}{Based on observations collected with European Southern
Observatory telescopes at La Silla, Chile}

\author{Adriano Fontana}
\affil{Osservatorio Astronomico di Roma, Via dell' Osservatorio 2,
        I--00040, Monteporzio, Italy}
\author{Sandro D'Odorico}
\affil{ESO, Garching bei M\"unchen,
        Karl-Schwarzschildstr. 2, D--85748}
\author{Emanuele Giallongo}
\affil{Osservatorio Astronomico di Roma, Via dell' Osservatorio 2,
        I--00040, Monteporzio, Italy}
\author{Stefano Cristiani}
\affil{Dipartimento di Astronomia dell' Universit\`a, 35122
	Padova, Italy}
\author{Guy Monnet}
\affil{ESO, Garching bei M\"unchen,
        Karl-Schwarzschildstr. 2, D--85748}
\author{Patrick Petitjean}
\affil{Istitut d'Astrophysique de Paris - CNRS, 98 Boulevard
Arago, F75014 Paris, France}

\date{11/12/97 ; 12/10/97} 
\begin{abstract} 
We present here the spectrum of the galaxy companion to the $z=4.7$
quasar BR1202-0725, in the optical range $6000 -
9000$\AA, corresponding to $1050 - 1580$\AA~ rest--frame.  We
detect a strong  \lya emission line at $z=4.702$, with an integrated
flux of $2\times10^{-16}$ergs cm$^{-2}$s$^{-1}$, and  a UV
continuum longward of the \lya emission at a flux level of
$\simeq 3\times10^{-19}$ergs cm$^{-2}$s$^{-1}$ \AA$^{-1}$. We fail to detect
any CIV$_{1550}$ emission with a $3 \sigma$ upper limit of
$3\times10^{-17}$ergs cm$^{-2}$s$^{-1} $.  We show that the ratio
between \lya and continuum intensity and the absence of a strong CIV
emission imply that the UV continuum radiation is the result of an
intense star--formation activity rather than of a reprocessing of the QSO
flux. The total estimated SFR of this $z=4.7$ star--forming region is $\sim
15-54$ M$_{\odot}$ yr$^{-1}$, depending on the IMF and the metallicity.

The present data suggest that the \lya emission has a velocity
and spatial structure, with possible velocity differences 
of $\simeq 500$ \kms on scales of few kpc. 
These velocity patterns may be a signature of collapsing or merging
phenomena in the QSO and its environment, as expected from current
models of galaxy formation at high $z$.

\end{abstract}

\section{Introduction}

In recent years, the association between low redshift quasars and
galaxies has been placed on firm observational grounds. HST and
ground--based surveys have shown that most, if not all, the $z<1$ QSOs
lie in host galaxy that often show clear sign of morphological
distortions and are preferentially located in overdense environments
(see e.g. Bahcall et al 1997 and references therein), suggesting
that the QSO activity and
evolution at low and intermediate redshifts are driven by interaction
between galaxies (Barnes and Hernquist 1991; Cavaliere and Vittorini
1998).

The QSO evolution at $z>>1$ is expected to be more closely related to
the formation of massive galaxies in overdense regions (Haenelt and
Rees 1991), and high redshift QSOs have long been thought to mark
preferential site of galaxy formation.  Now that viable techniques for
the selection of star--forming galaxies at $z\geq 3$ have been
envisaged (Steidel et al 1995, 1996; Madau 1995, Lanzetta, Yahil,
Fernandez--Soto 1996; Giallongo et al 1996; Hu and McMahon 1996), the
connection between QSO activity and early galaxy formation may be
investigated directly at these high $z$.  Few examples of \lya
emitting objects have already been found around $z\simeq 2.5$ QSOs
(Pascarelle et al 1996, Warren and M\o ller 1996), although \lya
eimitting companions are not common around radio quiet QSOs (Hu et al
1991).

Among the highest redshift QSOs, several investigations have recently
been focused on the radio--quiet BR1202-0725 (McMahon et al 1994), at
an estimated redshift of 4.694 (Storrie-Lombardi et al. 1996).
Fontana et al (1996, hereafter FCDGS) carried out a deep photometric
study of the field around the QSO in the B,V,Gunn$r$ and I bands. They
detected a clearly distinct nebulous object (hereafter referred to as
BR1202-07 B) of I magnitude $m_I = 24.1$, located 2.2 arcsec NW of the
QSO. They concluded that its colors can be
reconciled only with a galaxy at  $4.4 \leq z \leq 4.7$
where star formation has been going on for less than $10^{8}$ yr, and
derived a star formation rate $\sim 30$ M$_{\odot}$ yr$^{-1}$ (for a
IMF with slope $x=-1.5$ and solar metallicity).

A strong Ly$\alpha$ emission from this object at the same redshift as
the QSO has been detected in narrow band imaging by Hu et al.  (1996a)
and in imaging spectroscopy by Petitjean et al. (1996).  As shown by
Petitjean et al (1996), these results imply that the object can be
either a starburst proto--galaxy or a nebular cloud photoionized by
the QSO where part of the continuum is due to the reflection of the QSO
light by dust.

Excess millimeter and submillimeter emission consistent with dust has
been detected from this quasar (Isaak et al 1994; McMahon et al.
1994), and more recently, CO emission at $z=4.7$ has been found close
to the quasar and the companion (Omont et al 1996).  In the light of this
new evidence Hu et al (1996b) hint that most likely BR1202- 07B is a
neighboring galaxy merging with the QSO host galaxy and producing
enough extended gas to form a \lya companion ionized by the QSO.

We present here a new spectrum of the companion to BR1202-0725
in the range $\lambda\lambda 6000-
9000\AA $  and use it to rediscuss the different hypothesis
on the nature of this object.

\section {The Data}

\subsection{Observations and data reduction}

The imager-spectrograph EMMI at the ESO--NTT was used remotely from
Garching with ESO grism \#1 to obtain 4 exposure (1h each) during two
clear nights on January 29-31, 1996, with the seeing varying from 0.7
to 1.1 arcsec FWHM.  The CCD was binned $2\times1$ in the direction of
the dispersion giving a scale of 5.7\AA~/pixel and 0.27''/pixel in the
direction of the dispersion and perpendicular to it, respectively.
The long slit of 1.5'' width was centered on the QSO at a PA of
$-42^{\circ}$.  Observations of the standard star Feige 67 were
obtained on both nights in the same configuration.  The most
straightforward data reduction steps (bias removal, flat--fielding,
cosmic ray detection and flagging, wavelength calibration) were
performed with the standard MIDAS facilities.  The spectral resolution
as measured from the Argon lines of the calibration lamp is 15~\AA~ at
7200~\AA.

We have developed a specific software to perform the most critical
issue in the data reduction, which is the subtraction of the
contribution of the underlying spectrum of QSO from the one of the
galaxy.  After several tests, the PSF of the QSO has been fitted only
on the side where the companion is located (i.e. northwest) as the sum
of a Gaussian and of Lorentian function. Most of their parameter
(namely $\sigma_G$, $\sigma_L$, $\Gamma$, the common center and the
ratio of their intensity) were allowed to vary smoothly along the
dispersion to follow the variation of the PSF with wavelength.  The
other free parameters (i.e. the sky level and the QSO intensity) were
simultaneously obtained by $\chi^2$ minimization at each CCD column
(i.e. in the direction orthogonal to the dispersion). The pixels
contaminated by the emission from the companion were not used in the
fitting process. This procedure was verified to remove satisfactorily
the QSO tail on the other side (i.e. southeast) of the QSO, at the
distance corresponding to BR1202-07B.  This procedure was repeated on
each frame, and the QSO--subtracted 2--D frames were then wavelength
calibrated and summed.  Finally, the spectrum was extracted with a
1.65" wide window, starting at a distance of 1.1'' from the QSO (the
highest resolution image by FCDGS shows that the resulting aperture of
$1.65$''$ \times 1.5$'' does encompass the bulk of both the continuum
and emission line regions), flux calibrated and extinction--corrected.
The main source of uncertainity in the continuum level comes from the
accuracy of the QSO subtraction.  We estimate that for the
uncertainity in the continuum flux is $\leq 30 \%$.

\subsection{The spectrum of a $z=4.7$ galaxy}

Figure 1 shows the flux calibrated spectrum of 1202-07B.  The most
prominent feature is the \lya emission line centered at 6932~\AA,
corresponding to a redshift $z=4.702$.  The measured FWHM is
28.8~\AA~($1250$ km s$^{-1}$), corresponding to 24.6~\AA~($1050$ km
s$^{-1}$) when the instrumental resolution is quadratically subtracted.
We measure an integrated Ly $\alpha$ flux from BR1202-07B of
$2\times10^{-16}$ergs cm$^{-2}$s$^{-1}$, in agreement with the FCDGS
estimate from photometric data and with Hu et al 1996a and Petitjean
et al 1996

At the expected position of the CIV$_{1550}$, no emission line is
detected above the noise level. The resulting $3 \sigma$ upper limit
on the integrated flux is $\sim 3\times 10^{-17}$ergs
cm$^{-2}$s$^{-1}$, assuming the same FWHM of the \lya emission
line. Although the upper limit for the CIV emission line does not
exclude the presence of a faint AGN, it is about $7$ times lower
than the flux needed to explain the I band flux in the nebular
emission model (see fig. 4 of Petitjean et al 1996).  Furthermore, the
observed continuum level of $3\times10^{-19}$ergs cm$^{-2}$s$^{-1}$
\AA$^{-1}$ is consistent with that expected for the spectrum of a
star--forming galaxy with $m_I = 24.1$ (see FCDGS, figure 2).  In the
next section we'll show that these results imply that the continuum
emission from BR1202-07B is the result of a star--forming activity and
not the nebular reprocessing of the QSO emissivity.

A close scrutiny of the \lya emission from BR1202-07B reveals other
interesting clues on the nature of this object. Fig. 2 shows the 2-D
\lya emission after subtraction of the QSO.  Both the raw pixels and
the superimposed contours levels suggest that the \lya emission is
structured. First of all, a tail of \lya emission extends toward the
QSO, as shown also in high resolution narrow--band images (Hu et al
1996b). Furthermore, the 2-D shape of the central \lya emission appears
to be ``tilted'', with larger velocity shifts with respect to the
QSO systemic velocity at increasing distances from the QSO.
 In an attempt to quantify this effect, we have computed, at
each velocity bin, the distance of the peak of the \lya emission, by
fitting a 1-D gaussian to any column: the line connecting these points
is also shown in fig 2. 

Although the S/N and the spectral and spatial resolution are limited,
we note that this tilt, that is detected in each single
exposure, and clearer in those with better seeing, is not likely to
arise from a poor QSO subtraction. The
total integrated flux of the QSO tail is indeed about 20\% of the \lya flux,
and we estimate that the uncertainity on its subtraction is reduced to
few percent of the \lya flux.  We further note that since the \lya emission
from the QSO is very broad, and doesn't change
appreciably in intensity in the wavelength range shown in fig 2 (see also
fig2 of Petitjean et 1996), its subtraction cannot result in the asymmetric
shape shown in fig. 2, as well as in the extension of the \lya
emission toward the QSO. Finally, the tilt is fairly insensitive to the
different procedures of QSO subtraction that we tested.

\section {Discussion}
The most straightforward interpretation of the spectrum of BR1202-07B
is that the UV continuum is produced by intense star formation at
$z=4.7$.  We have converted the observed flux (as obtained from the I
magnitude m$_I$ = 24.1) into SFR using the latest version of the
Bruzual and Charlot evolutionary models (Bruzual and Charlot 1998),
that allow for different IMF and metallicities. The values obtained
range from 15 (47) M$_{\odot}$ yr$^{-1}$ for a Salpeter IMF with low
metallicity $Z = 0.02 Z_{\odot}$ to 46 (144) M$_{\odot}$ yr$^{-1}$ for
a Scalo IMF with solar metallicity, in a $\Omega = 1 $ ($\Omega = 0$)
universe with $H_0=50$ km s$^{-1}$Mpc$^{-1}$. As discussed by FCDGS,
the K band detection places stringent constraints on the age of the
stellar population, which cannot be larger than $10^8$ yr, regardless
of metallicity and IMF. The \lya emission in starburst galaxies is
less straightforward to compute, since it depends on geometrical
effects and dust extinction. The total integrated \lya emission of
$4\times10^{43}$ erg s$^{-1}$ is about twice what expected from
scaling the Charlot and Fall (1991) computation of the \lya emission
to the observed SFR, in the case of a Salpeter IMF with solar
metallicity.  

Due to the proximity of the QSO, indeed, part or most of the observed
emission might result from nebular reprocessing of the QSO ionizing
flux. As shown by Petitjean et al 1996, the nebular spectrum is
dominated by line emission, and a model can be envisaged where all the
photometric properties can be ascribed to the coincidence of \lya, CIV
and OII emission lines in the $r$, I and K band respectively. However,
the CIV equivalent width is expected to be even larger than the \lya
one to compensate for the low efficiency of the I filter at the
corresponding wavelength, so that a 90\% depression of the \lya flux
is required in this case.  The upper limit on  the CIV emission line and the
detection of the continuum in the optical spectrum allow to reject
this  model with a high degree of confidence.

Finally, the flat observed $I-K\simeq 1$ (that samples the rest--frame
spectral shape between 1500~\AA~ and 4000~\AA) of BR1202-07B argues
against a contribution from the reflection of the QSO continuum on the
dusty CO-emitting companion (Omont et al 1996).  Indeed, the QSO is
much redder than its companion ($(I-K)_{QSO}=2$), and a further
reddening of about $1$ magnitude is expected from the albedo curve of
dust, as computed under standard assumptions for graphite--free grains
(Pei 1992).

While the continuum reflects intense star formation activity, the QSO may
be the dominant source of ionizing flux that is responsible for the
\lya emission. Following Warren and M\o ller (1996),
indeed, we obtain that the distance from the ionizing source is of the order
of $\simeq 70$kpc, well within the range allowed by the geometry of
the system ($13$ kpc are the minimum distance possible given the 2.2"
separation, $600$~kpc is the distance obtained if the redshift
difference is taken as cosmological).

In conclusion, the observed spectral properties of the companion to
BR1202-0725 are consistent only with a burst of star formation at
$z=4.7$, with the nearby quasar possibly providing an extra source of
ionization. 

The observed 2--D velocity profile of the \lya$\alpha$ emitter is an
indication that the dynamical state of the \lya emitting gas is out of
equilibrium, and that gravitational infall or tidal interaction may be
occurring in this object. The extension toward the QSO of the \lya
emission may imply infall toward the QSO host, if the galaxy is
physically located in front of the QSO, and infalling toward it, that
would result in the larger redshift measured. Otherwise, the companion
galaxy is located beyond the QSO and receiding from it: in this case,
a close interaction might have recently occurred between the two
objects, leading to the burst of star formation in the companion and
to a fuelling of the QSO central black hole (Heahnelt and Rees 1993,
Lake, Katz and Moore 1998). The extension of the Ly$\alpha$ emission
toward the QSO may be a relic of this interaction.  However, some
caution must be taken in drawing these conclusions since the \lya
emission line in the Steidel's sample of $z\simeq 3$ galaxies is -
when detected - generally redshifted by up to several hundred \kms
relative to the interstellar absorption lines (Pettini et al 1997).

If we assume that the
observed tilt in the 2--D \lya emission comes from bulk motion on kpc
scale within the emitting gas, the internal velocity range
($400-500$km s$^{-1}$, see Fig.~2) is rather large for rotationally
supported disks, but arises naturally during collapse and merging of
pre--galactic clumps on a central potential (Haehnelt, Steinmetz and
Rauch 1997).

Thus, BR1202--07B appears to be the highest redshift counterpart of
other \lya emitting galaxies found in the environment of high $z$ QSOs
(e.g. Steidel et al 1991, Warren and M\o ller 1996).  In addition to
it, other objects are detected in the environment of the QSO. A
fainter galaxy companion is also detected SW of the QSO in the $I$
frame from HST, and recently confirmed to be a \lya and[O\thinspace
II] emitter at $z=4.7$ (Hu et al 1996b). A CO emitting object (Omont
et al 1996) is detected at a different redshift $z=4.6915$ than the
\lya emission, corresponding to $\Delta v = 550$ \kms, and is aligned
with BR1202-07B and with the QSO itself. Their alignment is consistent
with recent models of galaxy formation, that show that structure
formation and merging occurs along 1--D or 2--D structures
(``filaments''), a feature common to other QSO companions at high
redshift (M\o ller and Warren 1997 and references therein).

All this evidence show that we are detecting a high redshift galaxy
still far from the dynamical equilibrium, and with a high
star--formation rate in the individual pre--galactic components. 
It remains to be assessed whether this situation is typical of  $\geq 4$ quasar.
 At $m_I = 24.1$, the rest--frame  UV luminosity of 
BR1202-07B is $L_{1500} = 3.1 \times 10^{41} h_70^{-2}$ erg s$^{-1}$ \AA$^{-1}$
(for $q_0=0.1$), about 2.35 times brighter than the "average" lyman-break 
galaxy at $z\simeq 3$  (Pettini et al 1997).  
The case of BR1202-07B, where the star--formation activity has been identified 
from its broad--band colors, shows that 
multicolor surveys as deep as $m_I \simeq 25$ and $m_B \simeq 28$ 
and with excellent image quality are required 
to investigate the environment of high--$z$ QSOs  and to test
the cosmological scenario where QSO activity at
high $z$ is directly related to the early stages of galaxy formation.

\paragraph{Acknowledgements.}We thank P. M\o ller for providing 
information prior to publication and F. Governato
for useful discussions. Partial support has been provided by the ASI
contract 95--RS--38 and EC contract ERB FMRX--CT96--086.

\clearpage

\figcaption{Spectrum of the \lya galaxy companion to 
BR1202-0725.
The spectrum has been boxcar smoothed outside the \lya emission for
illustrative purposes.}

\figcaption{the 2-D \lya emission from BR1202-07B, after
subtraction of the QSO. Intensity contours of the \lya emission line
are superimposed. Wavelengths have been translated into velocity shift
from the \lya emission of the QSO at $z=4.694$. Each pixel in the $y$
direction corresponds to 1.6 kpc at $z=4.7$.  The solid line connects
the peak of the \lya emission at any velocity bin}

\clearpage
\plotone{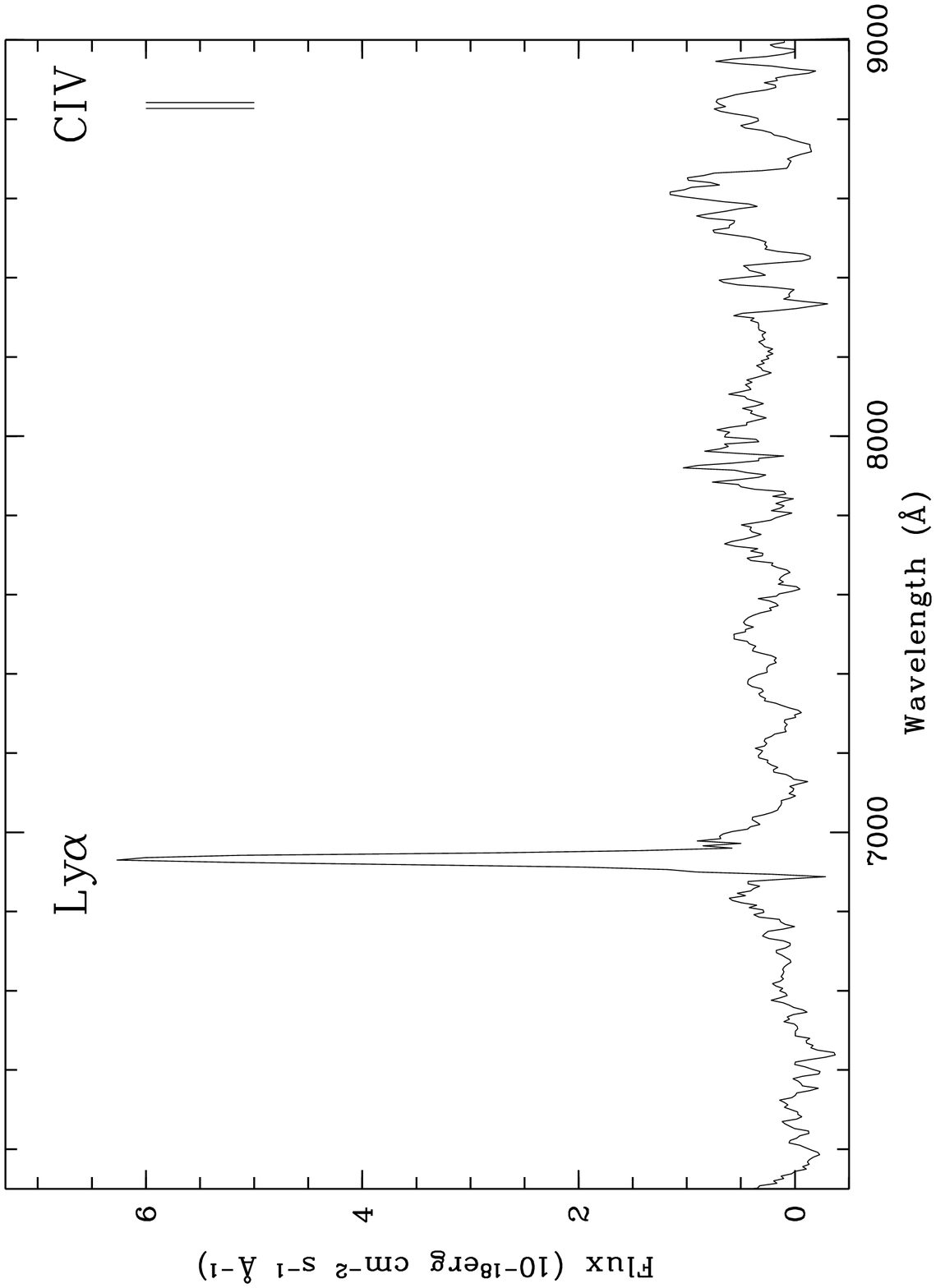}
\clearpage
\plotone{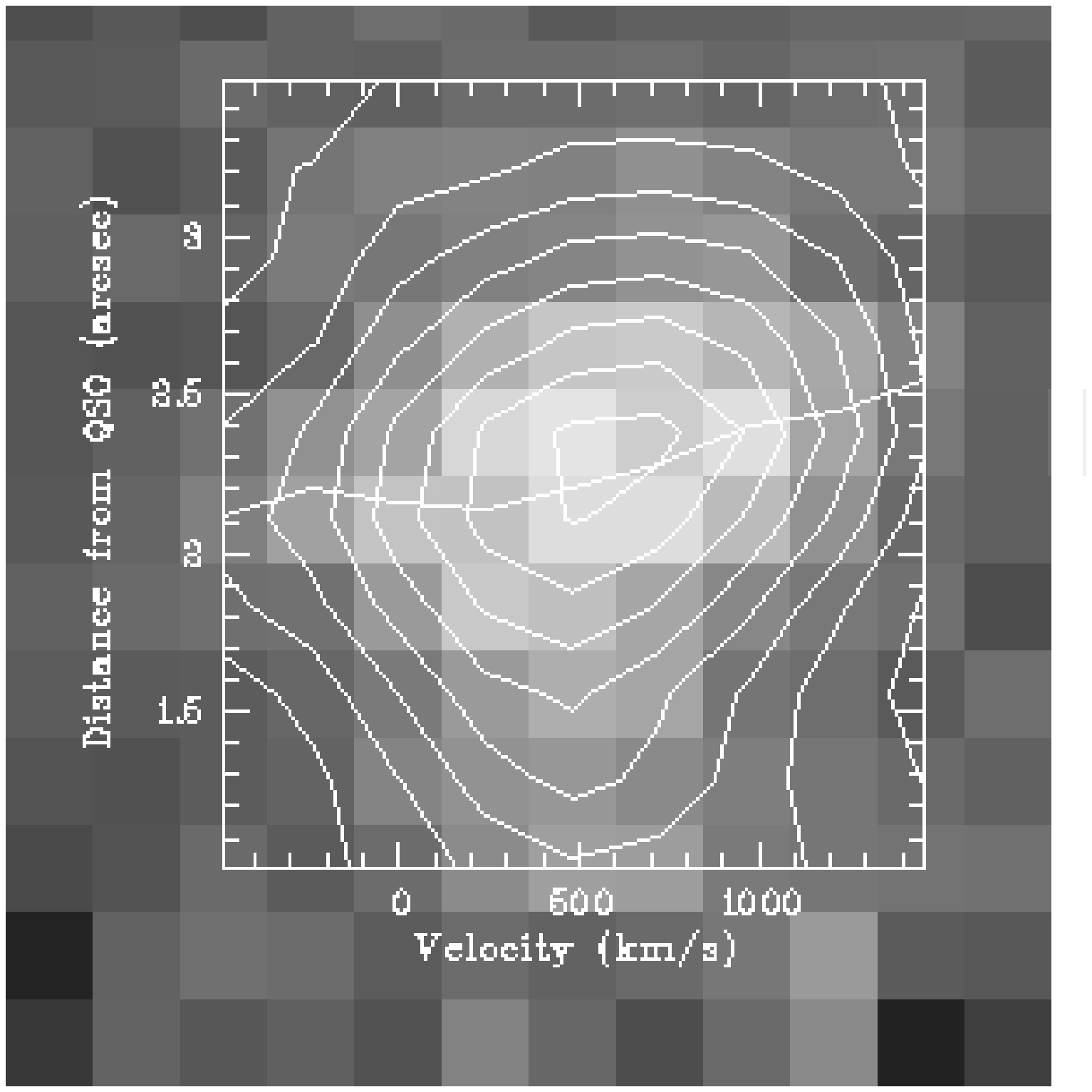}

\begin{thebibliography}{}


\bibitem[1986]{bes} Bahcall, J. N., Kirhakos, S., Saxe, D., Schneider, D. P., 
1997, ApJ, 479, 642

\bibitem[1986]{bes} Barnes, J.E., Hernquist, L.E., 1991, ApJL 370, 65

\bibitem[1986]{bes} Bruzual , G., Charlot, S., 1998 in preparation

\bibitem[1986]{bes} Cavaliere, A., Vittorini, V. 1998, in {\it The
Young Universe: Galaxy Formation and Evolution at Intermediate and
High Redshift''}, ed D'Odorico, Fontana, Giallongo

\bibitem[1986]{bes} Charlot, S., Fall, M., 1991, ApJ, 378, 471

\bibitem[1986]{bes} Fontana,\, A.,
Cristiani,\, S., D'Odorico,\, S., Giallongo,\, E., Savaglio,\, S. 1996,
MNRAS, 279, L27

\bibitem[1986]{bes} Giallongo,E., Charlot, S., Cristiani, S., D'Odorico S.,
Fontana A., 1996, in ``The VLT and the high redshift Universe'', p. 208
eds. Bergeron

\bibitem[1997]{bes} Haehnelt, M. G., Rees, M. J., 1993, MNRAS 263, 168

\bibitem[1997]{bes} Haehnelt, M. G., Steinmetz, M. and Rauch, M., 1997, ApJ submitted

\bibitem[1986]{bes} Hu, E.M., Songaila, A. Cowie, L.L., and Stockton, A., 1991,
ApJ, 368, 28

\bibitem[1986]{bes} Hu,\, E. M., McMahon,\, R. G.
1996, Nature, 382, 231

\bibitem[1986]{bes} Hu,\, E. M., McMahon,\, R. G., Egami,\, E.
1996a, ApJ, 459, L53

\bibitem[1986]{bes} Hu,\, E. M., McMahon,\, R. G., Egami,\, E.  1996b
in {\it ``HST and the high redshift Universe''}, ed. N.R. Tanvir,
A. Aragon--Salamanca \& J. Wall p. 91.

\bibitem[1986]{bes} Lake, G., Katz, N., Moore, B., 1998 ApJ in press,
astro--ph 9712012

\bibitem[1986]{bes} Lanzetta, K. M., Yahil, A., Fernandez--Soto, A.,
1996, Nature, 386, 759

\bibitem[1986]{bes} Isaak, K., McMahon, R.G., Hills, R.E. \& Withington, S., 1994, 
MNRAS, 269, L28.

\bibitem[1986]{bes} Madau, P., 1995, ApJ, 441, 18

\bibitem[1986]{bes} McMahon,\, R. G., Omont,\, A.,
Bergeron,\, J., Kreysa,\, E., Haslam,\, C. G. T. 1994, \+MNRAS, 267, L9

\bibitem[1986]{bes}  M\o ller, P., Warren, S.J., 1997, MNRAS submitted

\bibitem[1986]{bes}  Omont, A., Petitjean, P., Guilloteau, S., McMahon, R. G., Solomon, P. M., P\'econtal, E. 1996, Nature 382,428

\bibitem[1996]{pas} Pascarelle, S. M., Windhorst, R. A., Keel, W. C., Odewahn, S. C., 1996, Nature 383, 45

\bibitem[1986]{bes}  Pei, Y.C., 1992, ApJ, 395, 130

\bibitem[1986]{bes} Pettini, M., Steidel, C. C., Adelberger, K.,
Kellogg, M., Dickinson, M., Giavalisco, M., 1997 in ``Cosmic
Origings'', ASP Conference Series, eds. JM Shull, CE Woodward and HR
Thronson.

\bibitem[1986]{bes} Petitjean,\, P., P\'econtal,\, E.
Valls-Gabaud,\, D., Charlot,\, S. 1996, Nature, 380, 411

\bibitem[1986]{bes} Steidel,\, C. C., Sargent, W. L. W.,\& Dickinson,
M. 1991, AJ, 101, 1187

\bibitem[1986]{bes} Steidel,\, C. C., Pettini, M., Hamilton,\, D.,
1995, \+AJ, 110, 2519

\bibitem[1986]{bes} Steidel,\, C. C., Giavalisco,\, M., 
Pettini,\, M., Dickinson,\, M., Adelberger,\, K. L. 1996, \+ApJ, 462, L17

\bibitem[1986]{bes} Storrie-Lombardi,\, L. J., McMahon,\, R. G.,
Irwin,\, M. J., Hazard,\, C.  1996, ApJS, 468, 121

\bibitem[1986]{bes} Warren, S.J., M\o ller, P., 1996, A\&A, 311, 25

\end{thebibliography}
\end{document}